# Metasurface integrated Vertical Cavity Surface Emitting Lasers for programmable directional lasing emissions

Yi-Yang Xie[1], Pei-Nan Ni[2], Qiu-Hua Wang[1], Qiang Kan[3,*], Gauthier Briere[2], Pei-Pei Chen[4], Zhuang-Zhuang Zhao[1], Alexandre Delga[5], Hao-Ran Ren[6], Hong-Da Chen[3], Chen Xu[1,*], and Patrice Genevet[2,*]

[1] Key Laboratory of Optoelectronics Technology, Beijing University of Technology, Ministry of Education, Beijing 100124, China
[2] Université Côte d'Azur, CNRS, Centre de Recherche sur l'Hétéro-Epitaxie et ses Applications (CRHEA), Valbonne, 06560, France
[3] Institute of Semiconductor, Chinese Academy of Sciences, Beijing 100083, China
[4] National Centre for Nanoscience and Technology, Beijing 100190, China
[5] III-V Laboratory, Campus Polytechnique, 1 avenue Augustin Fresnel, Palaiseau, 91767, France
[6] Chair in Hybrid Nanosystems, Nanoinstitute Munich, Faculty of Physics, LudwigMaximilians-University Munich, Munich, 80539, Germany.

*correspondence to: kanqiang@semi.ac.cn, xuchen58@bjut.edu.cn, patrice.genevet@crhea.cnrs.fr

**Abstract**

Vertical cavity surface emitting lasers (VCSELs) have made indispensable contributions to the development of modern optoelectronic technologies. However, arbitrary beam shaping of VCSELs within a compact system still remains inaccessible till now. The emerging ultra-thin flat optical structures, namely metasurfaces, offer a powerful technique to manipulate electromagnetic fields with subwavelength spatial resolution. Here, we show that the monolithic integration of dielectric metasurfaces with VCSELs enables remarkable arbitrary control of the laser beam profiles, including self-collimation, Bessel and Vortex lasers, with high efficiency. Such wafer-level integration of metasurface through VCSELs-compatible technology simplifies the assembling process and preserves the high performance of the VCSELs. We envision that our approach can be implemented in various wide-field applications, such as optical fibre communications, laser printing, smartphones, optical sensing, face recognition, directional displays and ultra-compact light detection and ranging (LiDAR).

**Introduction**

Vertical cavity surface emitting lasers (VCSELs) have experienced a soaring development over the last 30 years, particularly after the demonstration of the first continuous-wave (CW) room-temperature device.[1-3] Their unique features such as low-power consumption, circular beam profile, wafer-level testing, large-scale two-dimensional array have made them the most versatile laser sources for a large number of applications, ranging from optical communications, to instrumentation, as well as laser manufacturing and sensing.[4-6] The ever-increasing application demands pose a longstanding challenge to further improve their performance while realizing precise beam control. In this context, the replacement of the top reflector with resonant structures and the incorporation of photonic crystal have been extensively employed to tune the emission and achieve high brightness. Meanwhile, considerable attention has been paid to improve the beam quality[7-9], for example, by preventing high-order transverse modes[10,11]. Although single-fundamental-mode laser can be realized by several methods, such as the reduced oxide aperture, the surface relief structure, the anti-resonant waveguide structure, *etc*, the typical divergence angle of the single mode lasers usually exceeds 10º, which needs to be further collimated.[12-14] To this end, both refractive and diffractive external micro-lenses have been incorporated into VCSELs to shape the beams. Pioneering efforts to integrate diffractive optical elements (DOEs), such as Dammann gratings and Fresnel lenses have led to DOE-VCSELs with superior beam quality operated at low deflection angles.[15-18] Today, artificial optical surfaces can outperform conventional DOE, in particular by improving the deflection performance at larger angles, to simultaneously control the phase and polarization states of light and even to produce complex field patterns, thus opening up new opportunities for wide-field laser applications. Metasurfaces, emerging as a new class of subwavelength-carved two-dimensional optical components[19-21], exhibit exceptional spectral and spatial controllability over the electromagnetic waves.[22-25] In comparison with conventional optical components, the unique planar configuration, large scale integration and complementary-metal-oxide-semiconductor processing compatibility of metasurfaces make them promising candidates for optoelectronic integration and perfectly suited for monolithic integration with VCSEL technology. Plasmonic metasurfaces have been integrated on the facet of the edge emitting quantum cascade lasers and VCSELs to improve the beam quality, light transmission and control polarization.[26-29] Recent high-index dielectric metasurfaces, predominantly considered for realistic applications, have demonstrated advantages of low absorption loss and the ability to engineer both the electric and the magnetic optical responses.[18,22-24] In this regard, high-contrast metasurface was integrated with VCSEL in replacement of top DBR mirror to modify the far-field emission patterns[30], and dielectric metasurface was integrated into intra-cavity to select a given vortex lasing emission by introducing a weak angular perturbation of light at the reflecting surface.[31] However, these integration approaches are highly intrusive as the integrations of metasurfaces modify the laser cavity, which not only affect the beam patterns but also change other lasing characteristics such as emission power and lasing wavelength, limiting their applications in arbitrary beam shaping. On the other hand, the static nature of dielectric metasurface does not allow arbitrary wavefront scanning or beam steering. The combination of the beam shaping metasurfaces with the programmability of laser beams of VCSELs chip, thanks to their unique two-dimensional array configurations, will be of great interest for future applications.

In this contribution, a wafer-level non-intrusive approach that solves the issues of arbitrary beam shaping VCSELs with programmable controllability is presented by sculpturing their emitting surfaces into metasurfaces. This monolithic approach provides an authentic new degree of freedom of beam shaping VCSELs without the risk of compromising their performance since the integrated metasurfaces operate as passive components, shaping the laser at the emission surface rather than altering the laser cavity. Featuring programmable laser emitting with fully-arbitrary beam profiles control, such lasers are named as Metasurface integrated Vertical Cavity Surface Emitting Lasers (MS-VCSELs). The efforts of maintaining the laser characteristics such as the operation mechanism, threshold, current distribution, lasing wavelength, high speed modulation, *etc*, have been considered, remaining fully compatible with existing state-of-the-art VCSELs technologies, including the wafer-level fabrication process, the standard packaging process, electrical injection solutions and theoretical analysis. Engaging metasurfaces with laser beams of well-defined wavelength and wavefronts, this approach benefits from the advantage of metasurface technology and resolves the critical diffraction issues of the most widespread laser systems to date. The arbitrary control of the wavefront at the wafer-level and programmability of MS-VCSELs would significantly promote their applications in optical fibre communications, laser printing applications, integrations into smartphone technologies, optical sensing, face recognition, directional display as well as high power applications such as ultra-compact light detection and ranging (LiDAR), benefiting from the fast developments of high power VCSELs.

**Design and fabrication principles**

MS-VCSELs are designed and fabricated into back-emitting configuration, as illustrated in Fig. 1a. The fabrication process flow are summarized in Supplementary Note 1. A large number of pairs of DBRs were adopted into the laser structure, which allows the operation of the MS-VCSELs under low CW injection currents without additional cooling apparatus (See Supplementary Note 2). Supplementary Fig. S3a and S3b show the lasing emission spectra under different CW currents, and reveal the single fundamental transverse mode operation of the lasers, which is achieved by reducing the oxide aperture to about 3 μm, as shown in the inset of Supplementary Fig. S3a. Since the design of meta-optics relies on spatial phase modulation of a given incoming wavefront, the single mode operation greatly simplifies the design of the metasurfaces. However, it is worth pointing out that the proposed wavefront shaping could as well compensate for complex multiple-modes laser operation (Supplementary Note 3 and Supplementary Fig. S4).

To define the size of the metasurfaces, the measured out-coupling intensity distribution at the back-side surface of a bare VCSEL was fitted with a Gaussian function. Beam diameter of about 86 μm was obtained, in good agreement with the estimated value given by the diffraction from the gain region considering an oxide aperture of 3μm. In order to maximize the interactions between the laser beam and the metasurface as well as to avoid the diffraction effects caused by the physical boundary of the metasurface, the diameter of the metasurface is fixed at a larger size of 200 µm (Fig. 1c), which will overlay the entire laser beam. In this design, centro-symmetric GaAs nanopillars of different diameters are used as polarization insensitive meta-atoms. It has been previously demonstrated that each nanopillar operates as an independent Fabry-Perot resonator with a low quality factor.[32,33] The phase and amplitude of the scattered light can be controlled by adjusting the pillar radius as determined by the finite difference time domain (FDTD Lumerical) simulations reported in Fig. 1b, in which $\theta$ represents the incident angle of the laser beam upon the metasurface. The height of the individual GaAs nanopillar was initially set at $h$=500 nm for the full wave simulations, and the deviation of the height caused by the fabrication inaccuracy was further studied (See Supplementary Note 4 and Supplementary Fig. S7). Increasing the nanopillar diameter results in better transversal confinement of light, increasing the effective refractive index, and thus larger transmitted phase delay.[34] Thanks to this simple and effective approach of phase control, the beam shaping can be realized by assembling nanopillars of different diameters at desired positions, according to specific designs.

**Collimated laser emissions with low beam divergence**

Collimating the emissions of the MS-VCSELs requires initial compensation for the beam divergence occurring during the propagation from the oxide aperture to the backside surface. This is achieved by considering, at first approximation, the field distribution emitted by a dipole placed at the gain region. Given that the oxide aperture is much smaller than the distance to the backside surface of wafer, we considered an out-going spherical wavefront with hyperbolic phase profile at the bottom surface. Note that this approximation neglects the filtering effect of the high finesse cavity. The latter could be accounted by doing full wave calculation of the propagation field. A desired phase delay $\phi_{collimator}(x,y)$ at a given position (*x*, *y*) with respect to the centre of the array, determined by

$$\phi_{collimator} = 2\pi - \frac{2\pi n}{\lambda}\left(\sqrt{x^2 + y^2 + f^2} - f\right) \qquad (1)$$

where *f* is the focal length, *n* is the refractive index of the substrate, is introduced to impart a compensating phase, collimating the incoming beam.

Considering the uncertainty of the nominal thickness of the GaAs substrate (625 μm +/- 25μm), different focal lengths in the range from 580 μm to 650 μm were adopted to optimize the collimation performance. Supplementary Fig. S9 illustrates the schematic of the characterization setup of the beam profiles. The measured beam intensity distributions along the propagation direction from the lasers with and without metasurfaces are shown in Fig. 2a, which are recorded from the adjacent lasers on the same chip under the same CW injection current (0.2 mA, *i.e.* slightly above the threshold current to avoid overheating the devices during measurements) for fair comparisons. It is found that the bare laser without metasurface diffracts the emitted light rapidly, while the emission characteristics of the MS-VCSELs have been remarkably modified by the integration of metasurfaces. Moreover, the optimal focal length can be obtained from the characterizations of the far-field patterns at *Z*=10 cm (Fig. 2b) and the divergence angles (Fig. 2c, Supplementary Note 5), respectively. It is found that the MS-VCSEL with $f$=630 μm exhibits the best collimation performance with a well-defined symmetric far-field pattern (Supplementary Fig. S13) and a very respectable divergence angle of about 0.83º, indicating that the actual thickness of the substrate is around 630 μm. In contrast, the bare VCSEL is highly divergent with a divergence angle about 30.4º (See Supplementary Note 5). Moreover, variation of the substrate thickness increases the divergence angle. The tolerance of substrate thickness for a chosen focal length design can be evaluated from Fig. 2c. For example, the substrate, which deviates +14 μm or -20 μm from the designed focal length, increases the divergence angle to 1.14º or 0.92 º, respectively. Note that such beam shaping approach is completely non-intrusive and does not modify the standard VCSELs architecture, while taking full advantages of the unique characteristic of metasurface in manipulating the electromagnetic fields to enable arbitrary control of wavefront with subwavelength spatial resolution. This was confirmed by comparing the *P-I-V* characteristics of the same laser before and after the fabrication of the collimating metasurface with focal length of 630 μm, as shown in Fig. 2d. It can be seen that the integration of metasurface barely affects the lasing characteristics of the devices, while it remarkably shapes the wavefront of the lasing beam. Furthermore, the transmission efficiency of the MS-VCSELs is estimated, based on laser characteristic without versus with metasurfaces, to be about 80% at the current of 0.2 mA, which is in good agreement with the high transmission design of the GaAs nanopillar building blocks. The transmission efficiency can be further improved by the optimization of the height of the nanopillars (See Supplementary Note 4). The collimation efficiency of the MS-VCSELs was further experimentally investigated at the same current, revealing high collimation efficiency of about 57% with a focal length of 630 µm (Supplementary Fig. S14). Further increasing the injection currents up to 7 mA leaded to redshifts of the lasing wavelength since no additional cooling was employed during the measurements (Supplementary Fig. S3b), which was supposed to have no effect on the collimation performance (See Supplementary Note 6). Moreover, it was experimentally confirmed that comparable collimation performance maintained under larger injection currents, indicating the validity of this approach at large optical powers (Supplementary Fig. S15). Despite that such collimation design was simplified for the MS-VCSELs under single fundamental mode operation, it can properly function even when the MS-VCSELs has multi-modes emissions (See Supplementary Note 3, Supplementary Fig. S4 and Supplementary Video 1). Additionally, the influence of the alignment accuracy during the metasurface integration has also been investigated (See Supplementary Note 7 and Supplementary Fig. S16).

**Examples of arbitrary beam profiles control**

Arbitrary wavefront shaping characteristics of MS-VCSELs can be readily implemented by further adding an additional phase response to the collimator phase profile. As an example, a non-diffracting zero-order ($J_0$) Bessel laser is demonstrated. The design is composed by superimposing the collimating phase delay ($\phi_{collimator}$) with an additional phase retardation to further deflect the collimated wave into an assembly of tilted plane waves with wave vectors distributed on a cone. GaAs nanopillars are thus assembled accordingly to add to the collimator phase $\phi_{collimator}$ with an additional conical phase profile

$$\phi_{axicon} = 2\pi - \frac{2\pi}{\lambda}\sqrt{x^2+y^2}\,NA, \qquad (2)$$

where $NA=\sin\theta$ is the numerical aperture. The total imparted phase, $\phi_{total} = \phi_{collimator} + \phi_{axicon}$, both compensates for diffraction and further decomposes the $J_0$ Bessel function as an ensemble of tilted plane waves propagating toward the axis of the laser with half angles given by $\theta = \sin^{-1}(k_{//}/k_0)$, where $k_{//}$ represents the transverse light momentum introduced by the metasurface, as illustrated in Fig. 3a. Figure 3b shows the beam profiles along the propagation direction under the current of 0.2 mA. The lasing beam of the device shows a well-defined, 160 μm long non-diffracting region along the propagation axis, which is in good agreement with the theoretical value predicted from the geometric optics, *i.e.* $\frac{D}{2\tan(\theta)}$=161 um, where D~86 um is the diameter of the incident Gaussian beam upon the metasurface. The intensity profile of the emitting beam can be well fitted by the corresponding $J_0$ Bessel function, as shown in Fig. 3c. The measured full width at half maximum (FWHM) of the Bessel beam is about 1.4 μm, which agrees well with its theoretical value of 1.35 μm calculated by $\text{FWHM}_{J_0} = \frac{0.358\lambda}{NA}$. According to the non-diffracting nature of the Bessel beam, its beam profile remains almost the same along the interference length, as evidenced in Supplementary Fig. S17, confirming the realization of zero-order Bessel MS-VCSELs. Likewise, the Gaussian beams of VCSELs can be readily converted into beams carrying specific orbital angular momentum (OAM) modes, simply by adding an additional spiral phase profile $\phi_{total} = \phi_{collimator} + \phi_{OAM}$ with $\phi_{OAM} = l\theta$, $l$ representing the topological charge of the vortex and $\theta$ the angle in the metasurface plane at $Z=0$. This method allows experimentally observing the build-up evolution of a laser vortex from the metasurface to the far field (See Supplementary Video 2 in which *l*=5 as an example).

**Ultra-compact programmable directional laser emissions**

The two-dimensional characteristics of MS-VCSELs and their state-of-the-art packaging techniques make them an ideal platform for ultra-compact programmable beam manipulation and generation applications, such as beam steering, multi-channel light sources on a chip, *etc*. As a proof of concept, a chip of 10*10 MS-VCSELs were fabricated with different deflection angles (Fig. 4a). Such configuration allows programming MS-VCSEL by MS-VCSEL to emit deflected beams at various angles. Fig. 4b and 4c show the intensity distributions of the beams along the propagation direction from the MS-VCSELs, of which metasurfaces are designed by introducing different phase gradients along *X*-axis, *i.e.* $\boldsymbol{\phi}_{total} = \boldsymbol{\phi}_{collimator} + \boldsymbol{\phi}_{deflector}$, where $\boldsymbol{\phi}_{deflector} = 2\pi - \frac{2\pi}{\lambda} x \sin \boldsymbol{\theta}$, $\theta$ is the deflecting angles varying from 0º to 15º (Supplementary Fig. S18a). The laser beams were deflected towards to *X*-axis at the desired angles, respectively, while remaining perpendicular to the *Y*-axis. Supplementary Fig. S18b show the radius of the beams deflected at different angles as a function of the propagation distance, which can be well-fitted into Gaussian functions. Furthermore, the deflected beams exhibit comparable far-field beam patterns with respect to the collimated beam (Supplementary Fig. S18c). The above facts reveal that well-collimated Gaussian beams with various deflection angles can be readily selected from the same MS-VCSELs chip. The deflection efficiency of the MS-VCSELs is estimated to be in the range from 40% to 60% at *Z*=5mm (Supplementary Fig. S19). The efficiency of the MS-VCSELs shows no significant drop as increasing deflection angles, indicating their capability of wide-angle operation. Moreover, similar beam deflection performance was confirmed under larger currents up to 5 mA (Supplementary Fig. S20). By operating individual laser of the chip, it becomes possible to steer laser beams in real-time with ultra-fast speed, thus enabling efficient programmable laser beam arrays for imaging and LiDAR applications. In addition, programmable beam generations on a chip as a compact multi-channels light source can be realized from the MS-VCSELs array. As an example, optical vortex beams with different topological charges, including $l=-2$, $l=-1$, $l=0$, $l=1$, $l=2$, were designed and experimentally demonstrated (see Supplementary Fig. S21). Despite that the integrated metasurfaces in this work are passive components, the integration technology developed in this work is fully compatible with the fast-growing field of electrical tuneable active metasurface, of which their functionalities can be tuned in an ultra-fast way by external stimuli, thus enabling dynamic wavefront shaping of VCSELs with high operation speed.

**Conclusions**

Metasurface integrated Vertical Cavity Surface Emitting Lasers (MS-VCSELs) featuring arbitrarily wavefront engineering of laser emissions at ultra-compact wafer-level are presented through a nonintrusive monolithic integration approach. Meta-atoms are assembled to generate collimated Gaussian beams, non-diffracting Bessel and OAM carrying lasers, respectively, to demonstrate their capabilities of arbitrary beam control. The feasibilities of MS-VCSELs in realizing programmable laser beam steering and multi-channel light sources generation are further explored. With respect to the widely reported wavefront engineering applications of metasurface for white light imaging, full-color hologram and so forth, where rigorous designs are required to correct chromatic aberrations for broadband operation as well as to solve the challenging issues of high efficient large-scale operations, the metasurfaces employed in this example are engaged with laser beams of well-defined wavelength and wavefronts within well-defined area, which feature the simplicity of monolithic integration of metasurfaces with high efficiency, representing an accessible and applicable solution for ultra-compact laser wavefront engineering. MS-VCSELs proposed in this work offer an additional degree of freedom to beam shape VCSELs, thus adding a powerful tool to the design of VCSELs. Moreover, their beam shaping properties can be integrated into the design of high performance lasers together with the other state-of-arts technologies, including high contrast grating VCSELs, photonic crystal VCSELs, benefiting from their remarkable laser emission characteristics, such as high power, single-mode operation and tunability.

**Data availability statement**

The data that support the plots within this paper and other findings of this study are available from the corresponding author upon reasonable request.

**Reference**

1. Koyama, F., Kinoshita, S., Iga, K. Room Temperature CW Operation of GaAs

**Acknowledgments**

We acknowledge the financial support from the National Key R&D Program of China (2018YFA0209000), National Natural Science Foundation of China (61604007, 61874145), Beijing Natural Science Foundation (4172009, 4182012). P. Ni, G. Briere and P. Genevet acknowledge the financial support from European Research Council (ERC) under the European Union's Horizon 2020 research and innovation program (grant agreement FLATLIGHT No 639109, and grant agreement i-LiDAR No 874986). The authors acknowledge the Nanofabrication Laboratory at National Centre for Nanoscience and Technology for sample fabrication. The authors thank Y. B. Gao, Y. H. Zhang and Z. H. Zhang for fruitful discussions.**Author contributions**

Y. Y. X., P. N. N. and P. G. conceived the idea and coordinated the experiment. H. D. C., C. X. and P. G. supervised the project. Y. Y. X., Q. H. W., Q. K., and P. P. C. carried out the fabrication, built the optical set-up, and performed the measurement. P. N. N., G. B., A. D., H. R. R. and P. G. conducted numerical simulations and supported the experiment with theoretical analysis. Y. Y. X., P. N. N., Q. K., Z. Z. Z., H. D. C., C. X. and P. G. performed data analysis. Y. Y. X., P. N. N., and P. G. wrote the manuscript draft. All authors participated in improving the final version of the manuscript.**Author Information**

The authors declare that they have no competing financial interests.

**Additional information**

**Supplementary information** is available in the online version of the paper.

**Reprints and permission information** is available online at www.nature.com/reprints.

**Correspondence and requests for materials** should be addressed to P. G.

**Methods**

Fabrication processing flow and characterization details are provided in the Supplementary Information.

**Figures:**

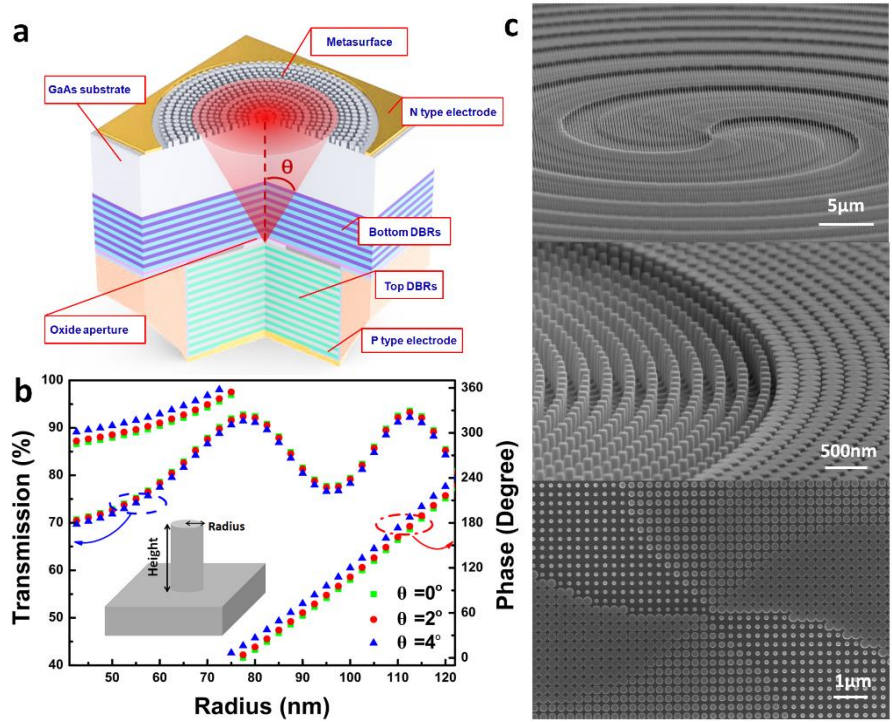

**Fig. 1 Design and fabrication principles. a,** Schematic of the MS-VCSELs, depicting the standard VCSEL structure and the beam shaping metasurface integrated at the back-side of the substrate, $\theta$ represents the incident angle of the laser beam upon the metasurface taking into account of the beam divergence and the substrate thickness. **b,** Calculated transmission and phase of GaAs nanopillars as a function of the radius for a laser emission wavelength of $\lambda=973$ nm upon different incident angles. The nanopillars were assembled in a subwavelength lattice considering the substrate refractive index to avoid spurious diffraction, fixing the distance between adjacent nanopillars at 260 nm. The inset represents a single nanopillar of the array. **c,** Scanning electron micrographs of a typical metasurface integrated with the MS-VCSEL. This example corresponds to a vortex emitting MS-VCSEL. Pillars with different diameters are disposed following the phase retardation profile to reshape the wavefront, compensating for the diffraction effect inside the wafer and addressing arbitrary wavefronts.

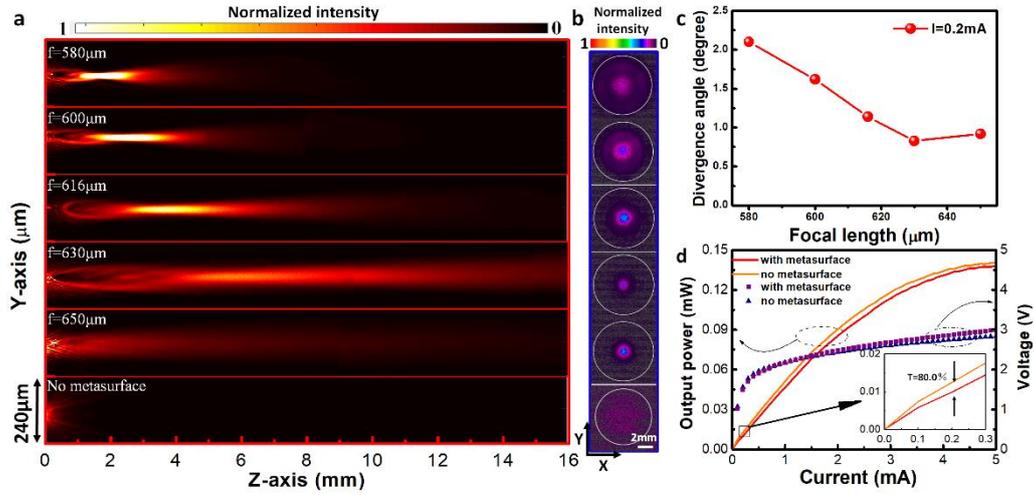

**Fig. 2 Performance of collimated lasers. a,** Performance of the collimating metasurfaces with different focal lengths. The optimal collimation performance is obtained when $f$=630 μm, as evidenced from the measured beam intensity distributions with and without metasurface. **b,** Far-filed beam patterns of the lasers at $Z$=10 cm, and **(c)** the divergence angle (evaluated using the beam's $1/e^2$ width) of the laser beam as a function of focal length. **d,** Comparison of the laser characteristics with and without metasurface, exhibiting similar *P-I-V* performance. **(d)** is of critical importance, indicating almost no degradation of the performance after the metasurface integration. The inset shows that the transmission efficiency of the integrated metasurface is estimated to be about 80% under the current of 0.2 mA.

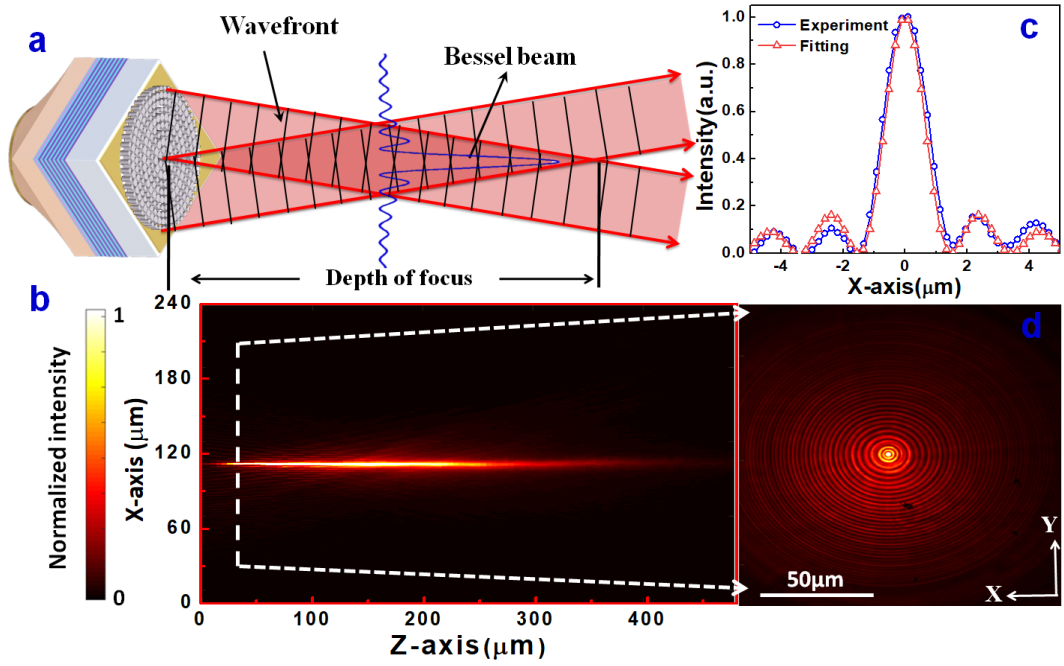

**Fig. 3 Generation of a zero-order Bessel laser. a,** Schematic diagram of the Bessel MS-VCSEL. **b,** Measured beam intensity profile along the propagation direction shows a non-diffracting emission region. **d,** The transverse plane intensity distribution at $Z=40$ μm as indicated by the dashed line in **(b)**. **c,** The measured intensity profile at $Z=40$ μm along $X$-axis agrees well with the zero-order Bessel function, confirming the generation of non-diffracting zero-order Bessel beam.

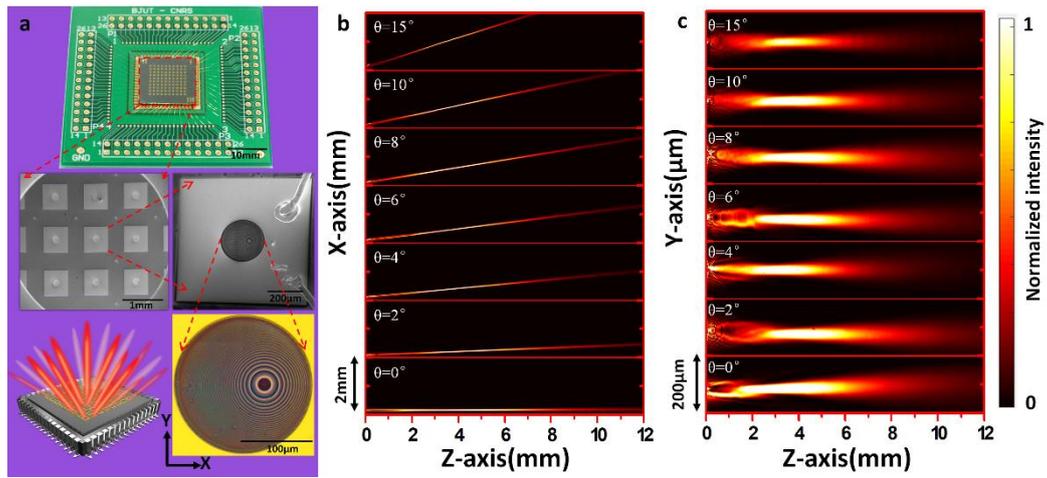

**Fig. 4 Programmable lasers array for wide-range dynamic beam steering. a,** Optical and SEM images of the array of 10*10 MS-VCSELs with different deflection angles mounted onto a PCB board. The inset shows the schematic of the MS-VCSELs chip with different deflection angles for wide-range dynamic beam steering applications. The measured beam intensity distributions in the *XZ*-plane **(b)** and in the *YZ*-plane **(c)** along the propagation direction with deflecting angles varying from 0º to 15º towards to *X*-axis.